\documentstyle[pramana,graphicx, epsf,floats]{ias}
\begin{document}

\title{Aging, rejuvenation and memory phenomena in spin glasses}

\author{V. Dupuis, F. Bert, J.-P. Bouchaud, J. Hammann, F. Ladieu, D. Parker and E. Vincent}
\address{Service de Physique de l'Etat Condens\'e (CNRS URA 2464), DSM/DRECAM, CEA Saclay,
91191 Gif sur Yvette - France}


\abstract{In this paper, we review several important features of
the out-of-equilibrium dynamics of spin glasses. Starting with the
simplest experiments, we discuss the scaling laws used to describe
the isothermal aging observed in spin glasses after a quench down
to the low temperature phase. We report in particular new results
on the sub-aging behaviour of spin glasses. We then discuss the
rejuvenation and memory effects observed when a spin glass is
submitted to temperature variations during aging, from the
point of view of both energy landscape pictures and of
real space pictures. We highlight the fact that both approaches
point out the necessity of hierarchical processes involved in
aging. Finally, we report an investigation of the effect of small
temperature variations on aging in spin glass samples with various
anisotropies which indicates that this hierarchy depends on the
spin anisotropy. }

\maketitle
\section{Introduction}
Spin glasses are magnetic systems in which the interactions are
disordered and frustrated, due to a random dilution of magnetic
ions. Below their glass temperature $T_g$, their dynamic response
to a small magnetic excitation is slow, and in addition depends on
the time spent below $T_g$ (`aging') \cite{Vincent:Sitges,Bouchaud:book}. In
contrast with structural and polymer glasses, aging in spin
glasses is hardly influenced by the cooling rate. On the contrary,
each cooling step leads to a restart of dissipation processes
(`rejuvenation'), while the memory of previous agings achieved at
different temperatures during cooling can be retrieved during 
re-heating \cite{Refregier,SaclayUpps}. It is possible to store independently the memory of
several isothermal agings, which corresponds to imprinting the
trace of various spin arrangements at well separated length scales
that are selected by temperature \cite{Bouchaud:PRB2002,Miyashita}.

In this paper, we first recall in section 2 some basic features of
isothermal aging in spin glasses. We discuss in particular the
scaling laws used to describe the response of a spin glass to a
field variation and report new results which raise the question of
the link between the cooling time and the sub-aging behaviour
observed in several spin glasses. We then examine in section 3 the
effect of temperature variations on aging which reveals striking
rejuvenation and memory effects and discuss these effects in terms
of energy landscape/real space pictures. Section 4 is devoted to a
comparison between an Ising spin glass and several Heisenberg spin
glasses with various degrees of random anisotropy, which shows
that the aging properties are markedly different in these systems,
the memory effect being more pronounced in the Heisenberg case \cite{Dupuis:PRB2001,Bert:PRL2004}.
These last results underline the role of anisotropy in the nature
of the spin-glass phase with Heisenberg spins and are discussed in
the context of a crossover expression for the relationship between
time and length scales in an aging spin glass.

\section{Isothermal aging in spin glasses}

\subsection{Basics features and scaling laws}

A large variety of systems exhibit slow glassy dynamics and aging
effects, i.e. history dependent properties. In all these systems,
the response to an external excitation depends on the time $t_w$
(waiting time) that the system has spent in the low temperature
phase before the excitation is applied. In most cases, the
response function can be typically expressed as the sum of a fast
stationary part and a slow aging contribution which roughly scales
as $t/t_w$ (Figure 1a) \cite{Vincent:Sitges,Bouchaud:book}:

\begin{equation}
R(t_w+t, t_w) = R_s(t) + R_a(t/t_w)
\end{equation}

\noindent where $t_w$ is the time elapsed at the measurement
temperature $T_m$ since the system has been cooled down below its
glass temperature $T_g$ , and $t$ is the time elapsed since the
excitation has been applied. This scaling means that the effective
relaxation time of the system below its freezing temperature is
mainly set by its age $t_w+t$. It is observed not only in the
response to a field variation but also in the {\it ac}
susceptibility (harmonic response) which obeys an equivalent
$\omega t_w$ scaling , where $\omega$ is the frequency of the {\it
ac} field \cite{Vincent:Sitges}.

\begin{figure}[htbp]
\begin{center}
\includegraphics[width=6.2cm]{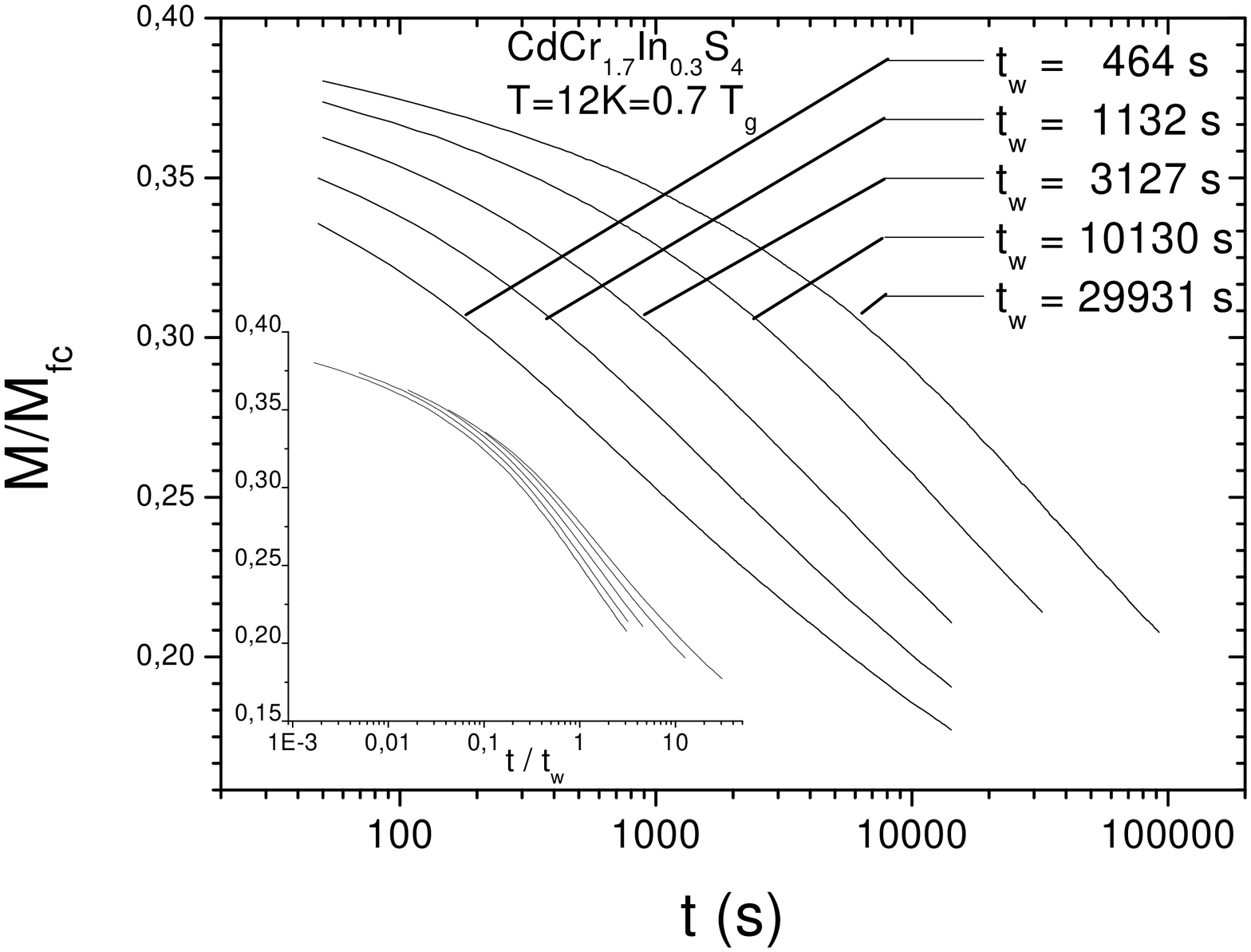}
\includegraphics[width=6.2cm]{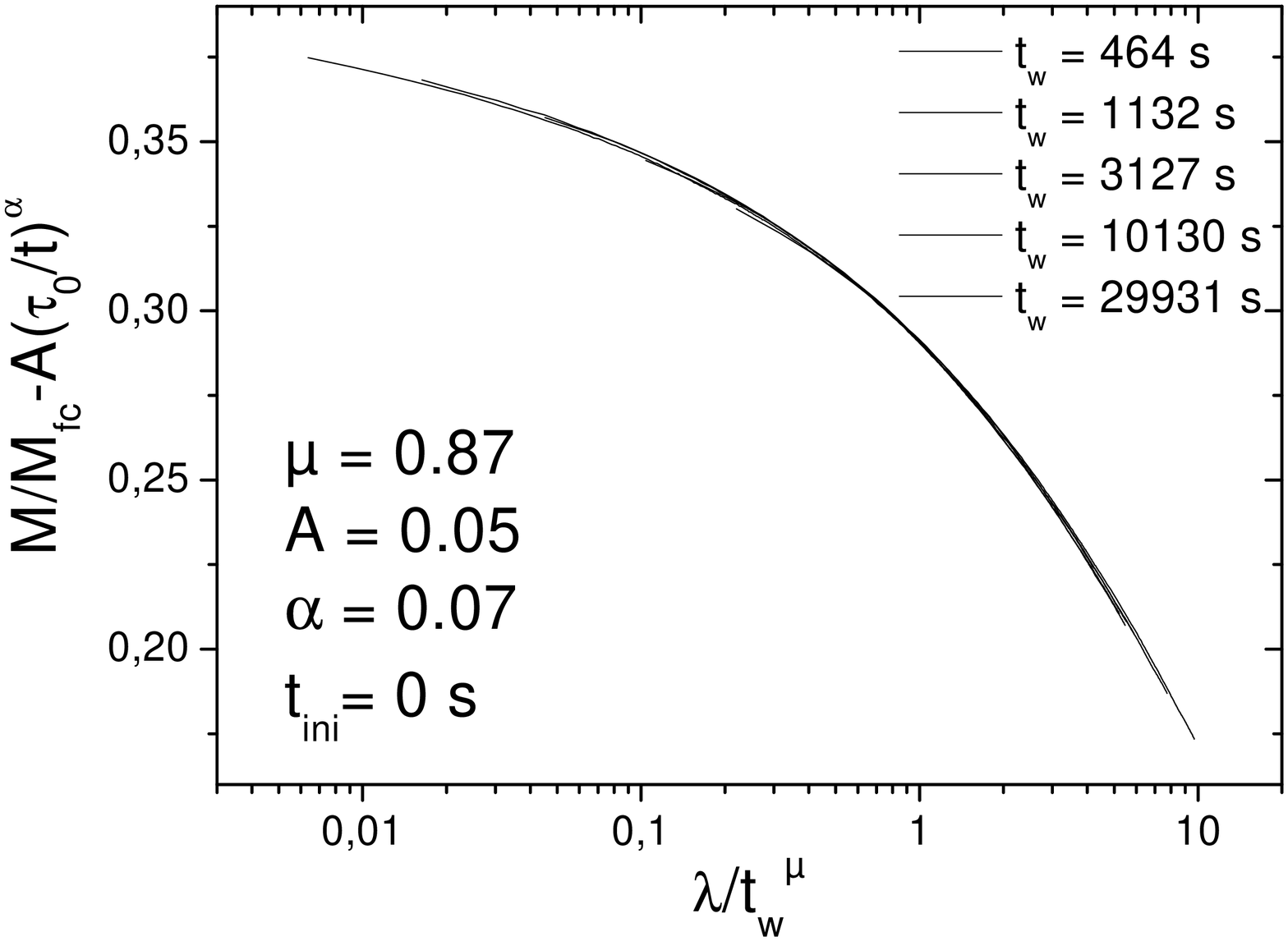}
\caption{(left) thermoremanent magnetization relaxations at 0.7 $T_g$
on the thiospinel CdCr$_{1.7}$In$_{0.3}$S$_4$ spin glass
($T_g=16.7\ K$); (right) scaling of the previous relaxation curves
(see text for details).} \label{fig:1}
\end{center}
\end{figure}

The $t/t_w$ scaling is however only approximate. It appears that a
more precise scaling can be obtained by assuming that the
effective relaxation time goes as the full age $(t+t_w)$ to a
certain power $\mu$  with $\mu < 1$ (Figure 1b). This leads to an
effective scaling variable $\xi = \lambda / t_w^\mu = 1/(1- \mu )[ (t+t_w)^{1-\mu}  -
t_w^{1-\mu}  ]$ (see \cite{Struik:book,Vincent:Sitges}). $\mu < 1$
means that the effective relaxation time grows more slowly than
the age of the system, this situation is thus called {\it
sub-aging}. Many examples of out-of-equilibrium systems such as
spin glasses, glassy polymers, microgel pastes and other soft
glassy materials, exhibit this sub-aging behaviour. In all these
cases, the exponent is smaller than 1, but remains close to 1 (of
order 0.8-0.9) in a rather large domain of temperatures
($0.4<T/T_g<0.9$). Such a behaviour is predicted in simple trap
models \cite{Rinn:PRL2000}, in some mean-field models of
spin-glasses \cite{Bouchaud:book} and in a recent extended droplet
model \cite{Yoshino:PRB2002,FH1988}.

\subsection{Intrinsic or extrinsic departure from full $t/t_w$ scaling}

The important question of whether the experimental value of $\mu$
smaller than 1 may not be due to experimental artefacts is however
far from being settled and has given rise to much interest
recently \cite{Rodriguez:PRL2003}. There are three main
experimental limitations which may have a crucial effect on the
results: the size limitations of the samples, the application of
finite excitations, the long time needed to cool the sample down
to the measurement temperature.

In a finite system composed of many subsystems (which could be
grains in polycristalline samples or powder), one can expect that
the ergodic times needed to explore the whole configuration space
of each subsystem remain also finite. In that case aging will be
interrupted at some ergodic time and the exponent $\mu$ will
decrease and eventually go to zero as the ergodic time is reached
in an increasing number of subsystems. This is a natural mechanism
for sub-aging which has been used in
\cite{Bouchaud:JPHYS1994,Joh:JPSJ2000} to estimate, from the
observed values of $\mu$, the cut-off of the barrier height
distribution or that of the size distribution of the considered
samples. However no experimental proof of the effect could yet be
given.

The amplitude of the applied magnetic field is obviously also very
important. Usually, only the minimum necessary excitations are
used in an aging investigation and care is taken to ensure the
response remains in the linear regime. However it is very
difficult to understand in which way and how much the system is
perturbed. A systematic study of the field effect was performed in
spin glasses \cite{Vincent:PRB1995} and showed that $\mu$ indeed
decreases as the excitation field is increased. However, it seems
to saturate at very low fields and stays at a value smaller than 1
as the field is reduced by large factors. In the case of the
thiospinel compound, $\mu$ saturates around 0.85 below 10 Gauss down to the smallest field probed ($10^{-3}$ Gauss)
\cite{Herisson}.

Recently some detailed investigations were devoted to the problem
of the cooling time and the importance of the initial conditions
in the measurement of the thermoremanent magnetisation (TRM) in
spin glasses
\cite{Zotev:PRB2003,Rodriguez:PRL2003,Nordblad:HFM2003}. Different
cooling procedures were used and their effects on the TRM were
considered. In the quoted investigations, three different cooling
protocols were used: (A) Direct cooling from above $T_g$ to the
measuring temperature $T_m$ (usual procedure); (B) Fast cooling
from above $T_g$ down to a temperature smaller than $T_m$ and
heating back up to $T_m$; (C) Cooling down to a temperature larger
than $T_m$ and waiting a certain time before the final cooling
step down to $T_m$.

Some differences have been found using these various initial
conditions. In \cite{Rodriguez:PRL2003} it was argued that a full
aging ($\mu=1$) could almost be obtained in a Cu:Mn$_{6\%}$
spin glass sample with protocol (B). In \cite{Nordblad:HFM2003},
the difference between the two protocols (A) and (B) in a
Ag:Mn$_{11\%}$ sample is only to reduce the initial age of the
system for the shorter cooling time (protocol (B)).

\subsection{Revisiting the link between cooling time and sub-aging}

In order to clarify the situation, we have performed new
experiments using protocols (A),(B),(C) described above, to check
the effects of different initial conditions on the scaling
behaviour of the response function of three well characterized spin
glasses.

In order to precisely estimate the real cooling times associated
with a given protocol, we recorded the field cooled magnetization,
which actually slightly depends on temperature and can thus be
used as a thermometer giving the true temperature of the sample.
In all cases, TRM measurements made just after the temperature
stabilization at $T_m$ show a maximum of dM/d(logt) at a time of
the order of the real cooling time. In other words the cooling
time in these experiments corresponds to some equivalent initial
waiting time $t_{ini}$ at $T_m$.

\begin{figure}[htbp]
\begin{center}
\includegraphics[width=6cm]{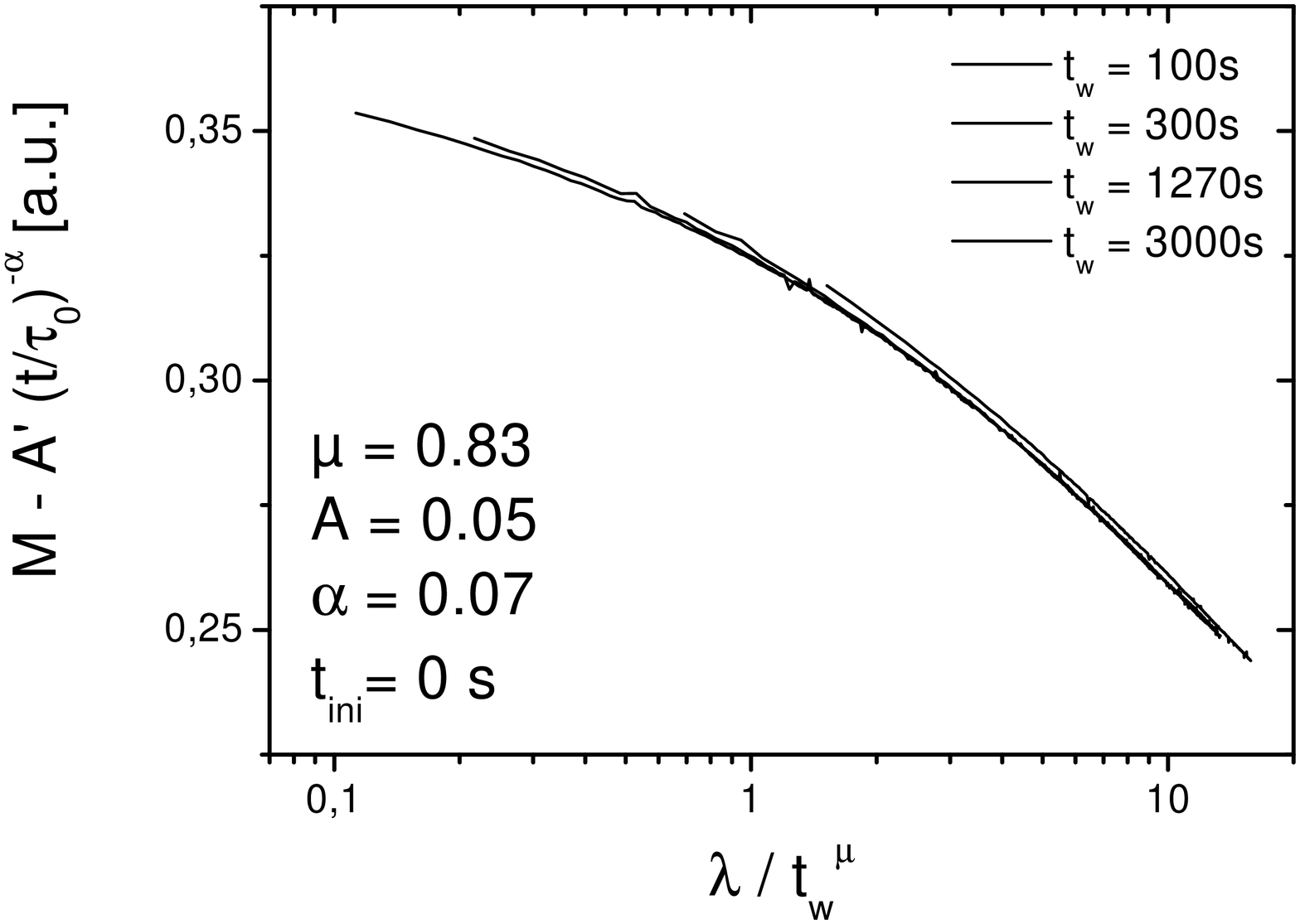}
\includegraphics[width=6cm]{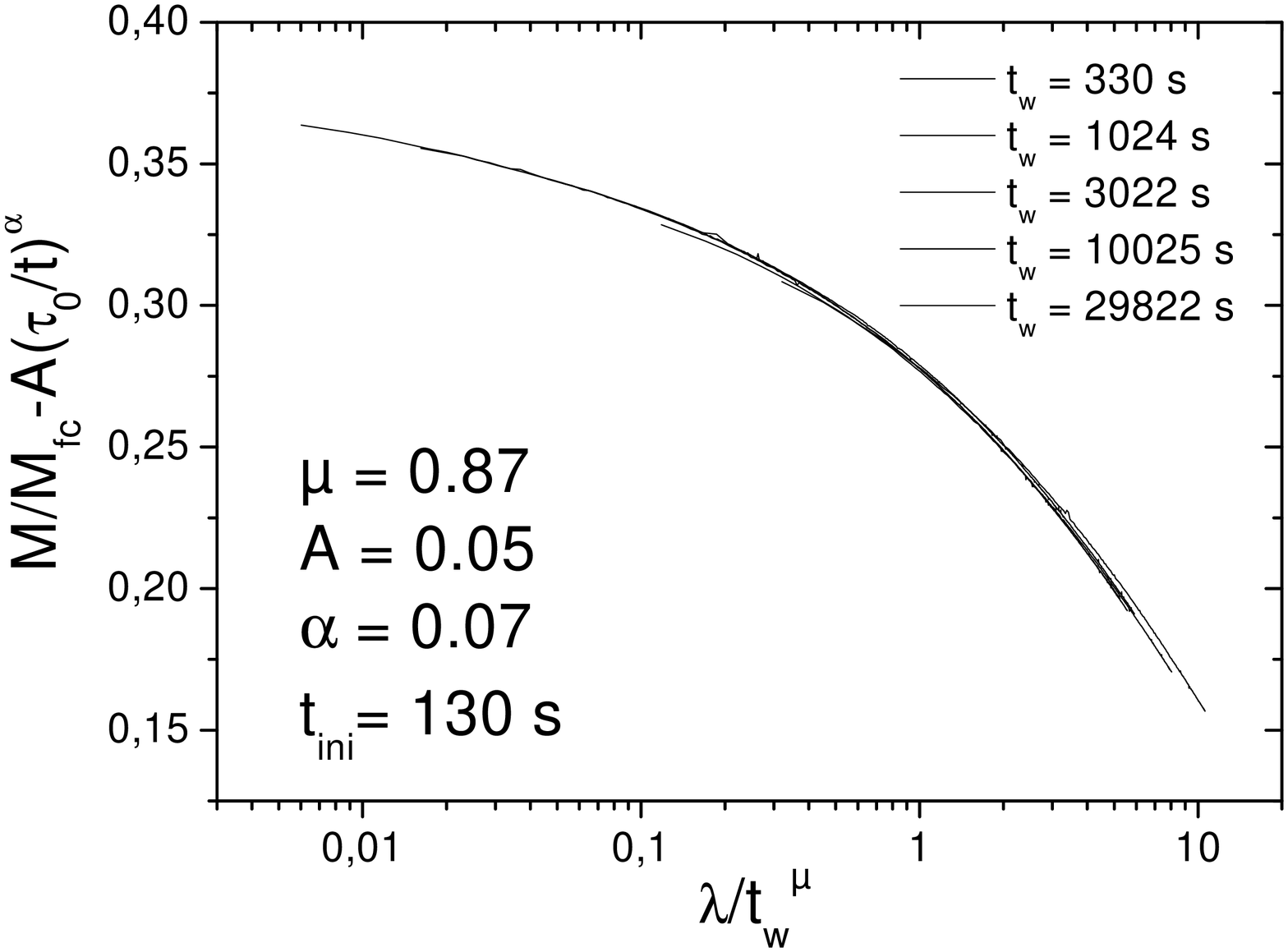}
\caption{Scaling of thermoremanent magnetization relaxations
measured on the CdCr$_{1.7}$In$_{0.3}$S$_4$ spin glass at $0.7\
T_g$. (left) fast cooling (protocol (B)) (for this set of data, the exact vertical scale is unknown, but within the error bars A' corresponds to A=0.05) ; (right) slow cooling (protocol
(C)) (see text for details).} \label{fig:2}
\end{center}
\end{figure}

Fig. \ref{fig:2}a shows a scaling of several TRM curves measured
according to protocol (B) for various waiting times with the 
Heisenberg-like thiospinel CdCr$_{1.7}$In$_{0.3}$S$_4$ spin glass.
The same sub-aging behaviour as in protocol (A) (Fig. \ref{fig:1}b)
can be observed with a slightly decreased value of the exponent
$\mu$ ($\mu=0.83$ here, compared to $\mu=0.87$ in protocol (A)).
However an experiment made following protocol (C), showed the same
result ($\mu=0.87$) as with the usual cooling protocol (A), even though the sample was kept for 1 hour at
$T/T_g=0.94$ before being cooled down to the measuring temperature
$T_m/T_g$ (Fig. \ref{fig:2}b). Other experiments on a Au:Fe$_{8\%}$
sample, another Heisenberg-like spin glass which is expected to
behave similarly as the thiospinel \cite{Bert:PRL2004}, do not
show any clear trend when slowing down the cooling procedure. The
results are compatible with the thiospinel results, though less
accurate due to a smaller signal, and point thus towards the same
tiny cooling effects, if any.

\begin{figure}[htbp]
\begin{center}
\includegraphics[width=6cm]{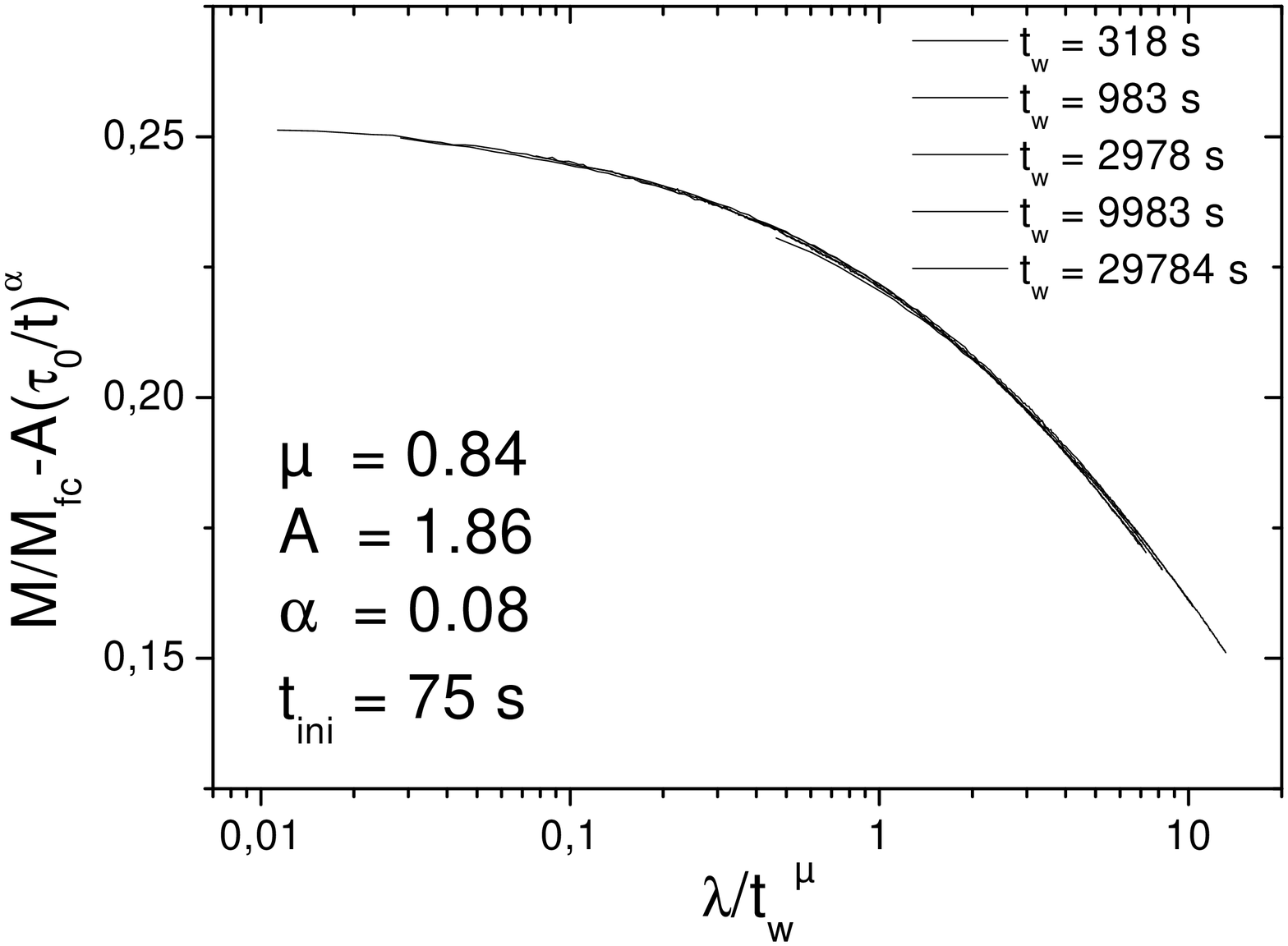}\includegraphics[width=6cm]{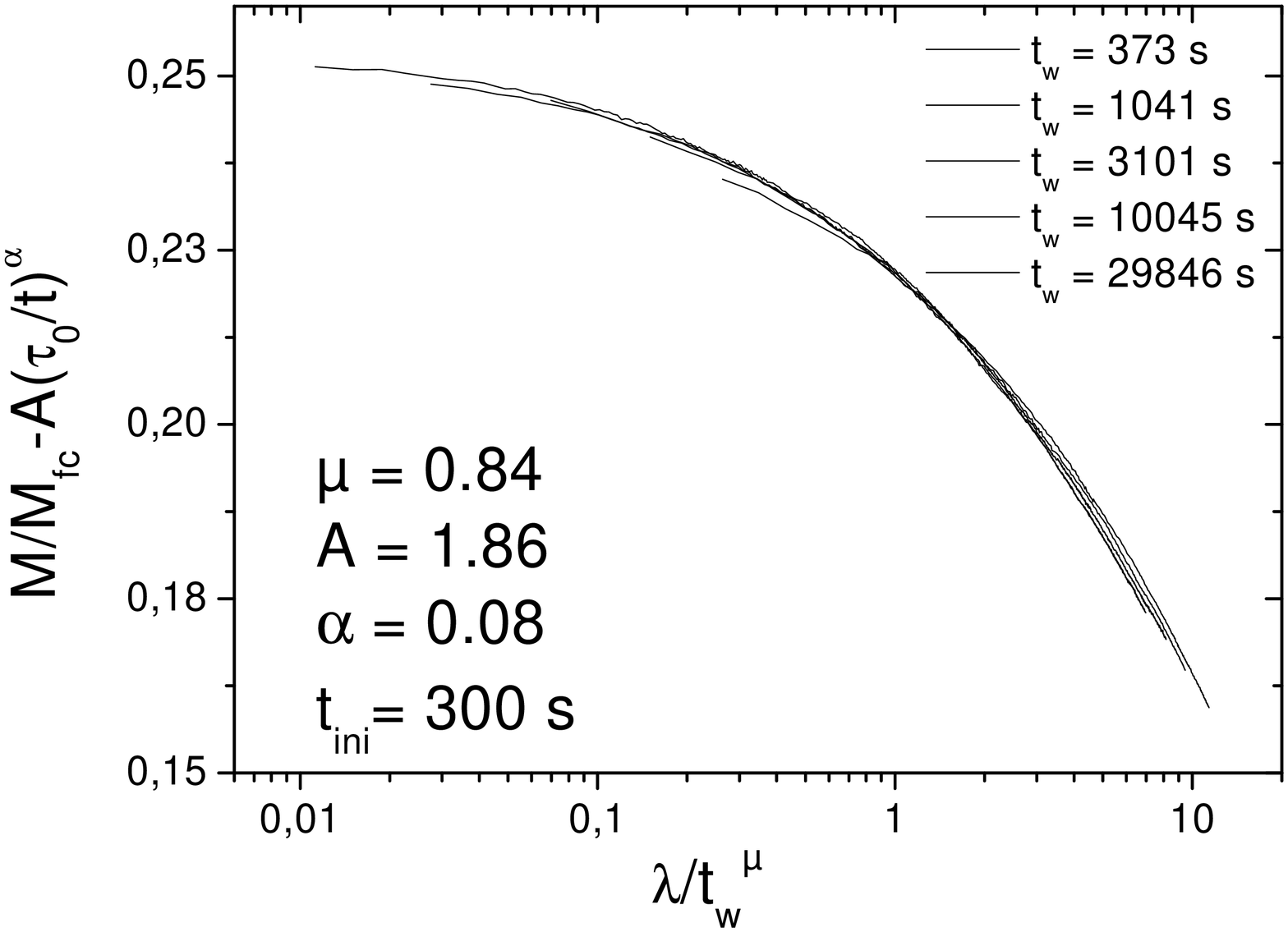}
\vskip 0.1cm\includegraphics[width=6cm]{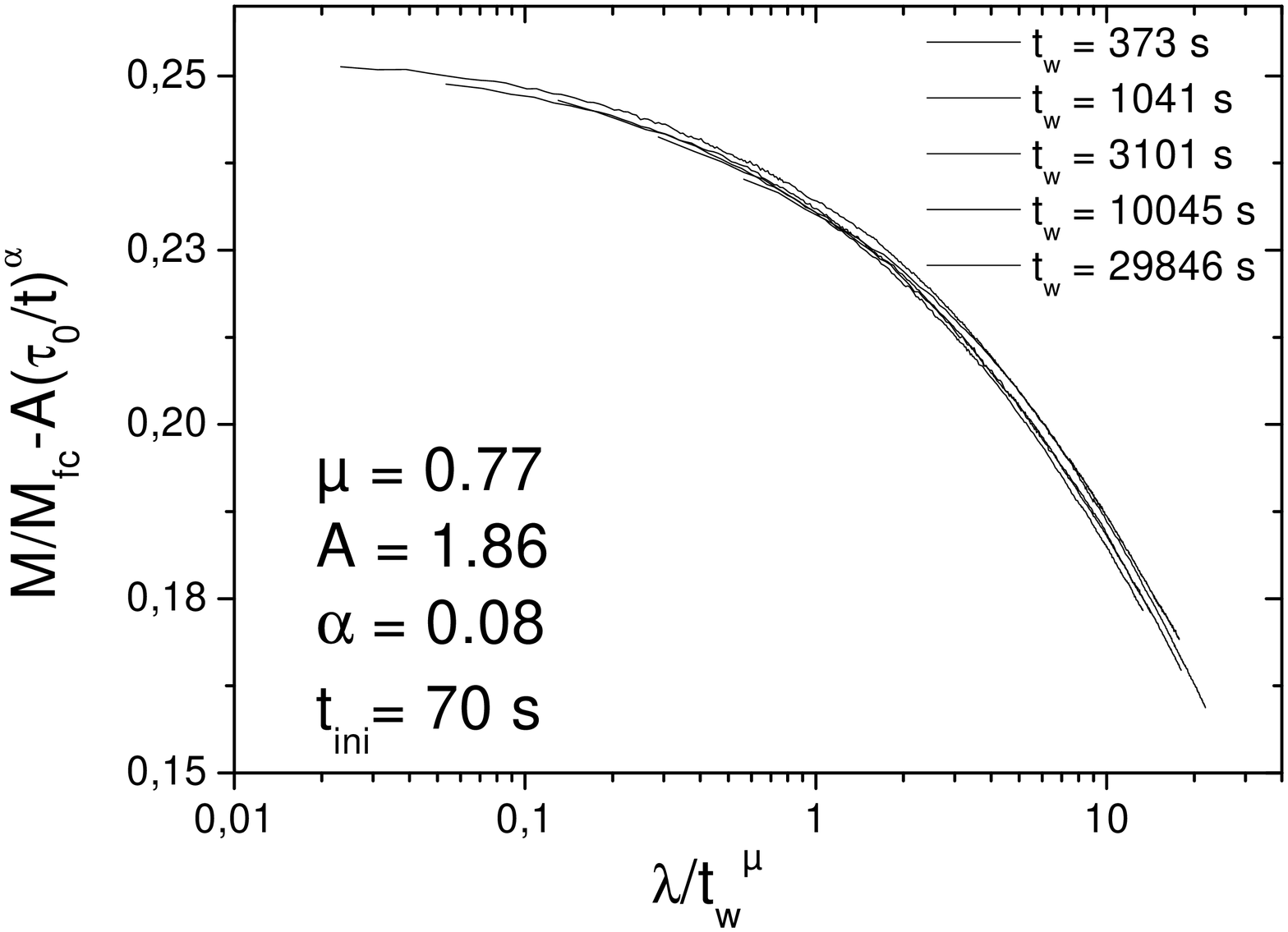}
\caption{Scaling of thermoremanent magnetization relaxations
measured on the Fe$_{0.5}$Mn$_{0.5}$TiO$_3$ Ising spin glass at
$0.7\ T_g$: (top left) fast cooling (protocol (B)) ; (top right) slow cooling
(protocol (C)), scaling with fixed $\mu$; (bottom) slow cooling (protocol (C), same data), 
but with  fixed $t_{ini}$.} \label{fig:2bis}
\end{center}
\end{figure}

We performed the same experiments on 
Fe$_{0.5}$Mn$_{0.5}$TiO$_3$, an Ising spin glass in which stronger
cooling effects can be expected \cite{Dupuis:PRB2001}. The scaling
of the TRM relaxation curves measured with fast (protocol (B)) and slow
(protocol (C)) coolings is presented in Fig. \ref{fig:2bis}. Fast
cooling (\ref{fig:2bis}a) yields $\mu=0.84$ and $t_{ini}=75\ s$,
with a good scaling. Slower cooling yields, as can be seen in
\ref{fig:2bis}b and \ref{fig:2bis}c, (i) a downgrading of the
scaling quality, as noted in \cite{Rodriguez:PRL2003} (ii) a
change in the scaling parameters which can be as well accounted
for by an increase of $t_{ini}$ or a decrease of $\mu$.

From this analysis, we conclude that there is no
evidence that $\mu$ would go to $1$ for a very fast cooling. Our
results are compatible with the idea of an `intrinsic' nature of the sub-aging
behaviour in spin glasses. In {\em numerical} simulations of (Ising) spin glasses, in which 
cooling can be instantaneous, it has been shown  
\cite{Berthier:PRB2002} that the effect of cooling rate on the value of $\mu$ 
very quickly saturates. $\mu$ was found to evolve from a value slightly {\it above} $1$ 
for infinite cooling rate to a value less than $1$ when the cooling rate is of the
order of $10^{-2}$ in microscopic units, and does not evolve significantly when the
cooling is slower. Since a cooling rate of $10^{-2}$ is still very much faster than
any experimentally achievable cooling rate, this scenario observed in numerical simulations may
be a plausible explanation for sub-aging in real materials. 

\section{Memory effects and hierarchy of states/length scales}

In the previous section, we discussed what happens when one simply
quenches a spin glass below its glass transition temperature $T_g$
and lets it evolve at constant temperature. Other protocols may be
imagined in which one can probe the effects of temperature
variations during aging and see, for example, how far heating slightly
an aging spin glass can help it to reach its equilibrium \cite{Refregier,SaclayUpps}.

\begin{figure}[htbp]
\epsfxsize=8cm \centerline{\epsfbox{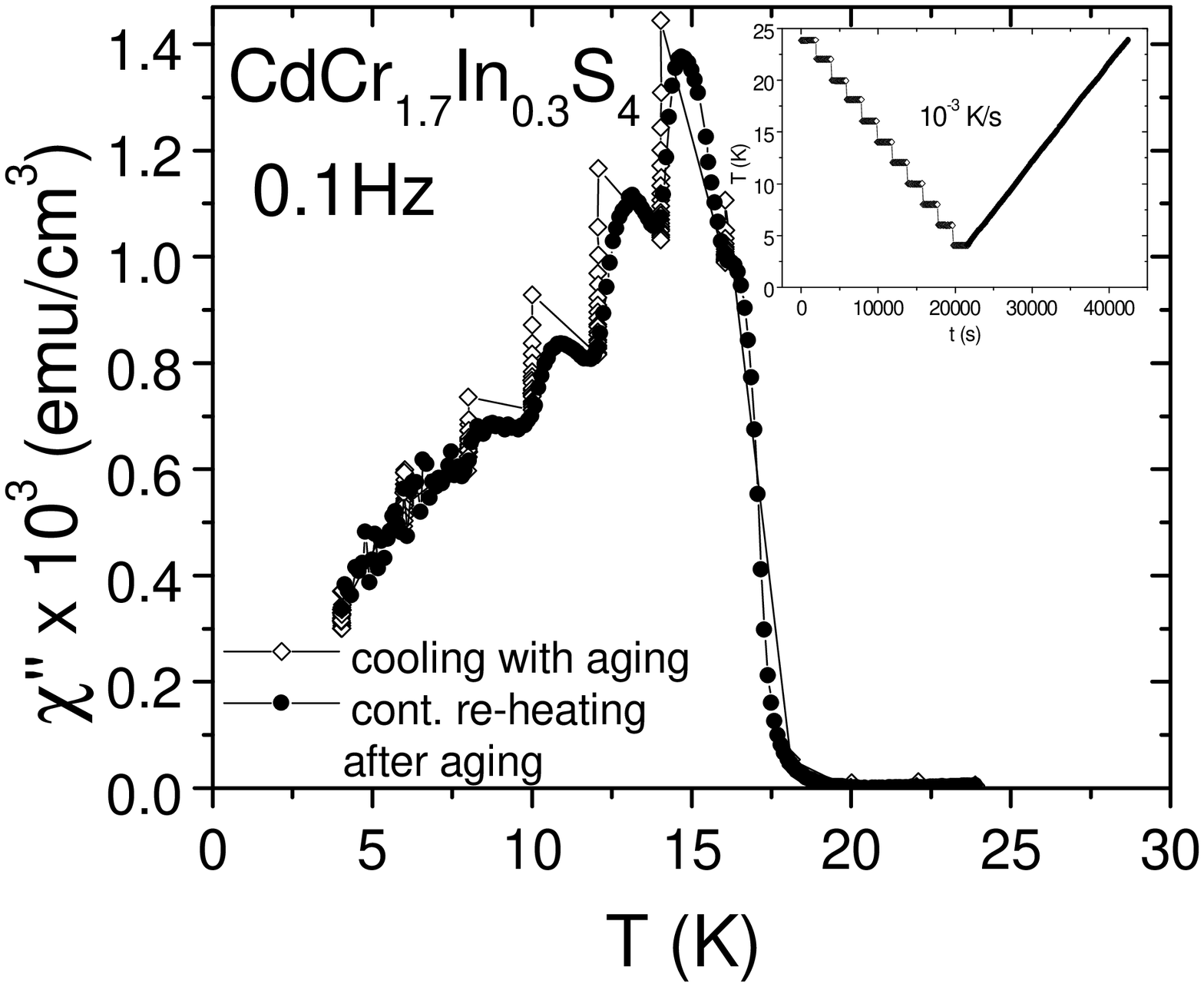}} \caption{Series
of dips imprinted on the ac susceptibility by successive stops at
different temperatures while the system is cooled.} \label{fig:3}
\end{figure}

Figure \ref{fig:3} shows the result of such a protocol on the
imaginary part $\chi''$ of the {\it ac} susceptibility measured at
$0.1\ Hz$ on the CdCr$_{1.7}$In$_{0.3}$S$_4$ spin glass. In this
experiment, the sample was step-cooled through its glass
transition temperature $T_g=16.7\ K$ down to $5\ K$ (each plateau
at constant temperature lasts approximatively 30 min) and reheated
then continuously up to above $T_g$. During each stop at constant
temperature, aging is directly evidenced as a downward relaxation
of $\chi''$. Looking carefully at the figure, we see that further
cooling seems to rejuvenate the system as the susceptibility rises
back: this is the {\it rejuvenation} effect. However, when heated
back continuously (full symbols), $\chi''$ shows marked dips at
the plateau temperatures where the sample was allowed to age,
revealing that the system has kept the {\it memory} of each stop
that occurred during the cooling.

\vskip 0.5cm\noindent {\bf Energy landscape pictures:} These
striking rejuvenation and memory effects can be interpreted in
terms of a hierarchical organization of the metastable states in
which the system evolves during its search for equilibrium \cite{Refregier,Vincent:Sitges}. This
picture, originally inspired from the mean field solution of the
SK spin glass model, was further developed in the past years in
several toy models such as the `traps on a tree' model \cite{Bouchaud:JPHYS1995} or the Sinai
model (in which the hierarchy is put by hand) (\cite{Sales}). On the other hand,
recent numerical simulations of the Generalized Random Energy Models
\cite{Sasaki}
(roughly equivalent to the `traps on a tree' model)
proved unambiguously that the idea of a hierarchy of metastable
states is able to capture the main features of experimentally observed aging.

\vskip 0.5cm\noindent{\bf Real space pictures:} While energy landscape pictures 
are illuminating, one cannot avoid trying to interpret the dynamics in an aging spin
glass in real space, that is in terms of spin configurations.
It is now widely accepted that aging in spin glasses corresponds
to a growth of a certain coherence length (which would be the
equivalent of a kind of out-of-equilibrium correlation length) \cite{FH1988,Bouchaud:PRB2002}. In
a ferromagnet, for example, this coherence length is simply the
size of the ferromagnetic domains and the aging observed after a
quench just reflects the growth of these ferro\-magnetic domains. In
spin glasses however, and contrary to ferromagnets where the
domain growth is directly related to cooling rate (the longer
the time spent at higher temperature, the larger the domains), aging is hardly influenced
by the cooling rate, as evidenced in Fig. \ref{fig:3} where each
cooling step seems to entirely rejuvenate the system.

A simple real space scenario which allows one to reconcile the
idea of a growing coherence length with the observed rejuvenation
and memory effects was discussed \cite{Bouchaud:PRB2002}. It was
noticed that a thermally activated dynamic over energy
barriers $B\sim \Upsilon(T)\ell^{\psi}$ that vanish at $T_g$ as
$\Upsilon(T)\sim(T-T_g)^{\psi\nu}$ ($B$ is the barrier to overcome
in order to flip a cluster of spins of typical size $\ell$)
naturally leads to a hierarchy of the length scales probed in
experiments. More precisely, and as described in
\cite{Bouchaud:PRB2002}, the temperature $T$ at which aging is
observed sets the length scale $\ell^*(T)$ at play in aging (often
called {\it active length scale}). Larger length scales
($\ell>\ell^*(T)$) are frozen while smaller length scales
($\ell<\ell^*(T)$) are fully equilibrated. Decreasing the
temperature from $T$ to $T-\Delta T$ reduces significantly this
typical length scale as $\ell^*(T-\Delta T)< \ell^*(T)$. This has
two important consequences. Firstly the structure previously
reached at length scale $\ell^*(T)$ freezes since the time needed
to reconform details of the structure at this length scale has
grown astronomically large. This will allow the frozen
structure to be retrieved by heating back the system (memory effect). Secondly,
the length scales $\ell$ such that $\ell^*(T-\Delta
T)<\ell<\ell^*(T)$, which were equilibrated at $T$, are no longer
equilibrated at $T-\Delta T$ due to the change with temperature of
the associated Boltzmann weights. They start to age giving rise to
the rejuvenation effect. This scenario, which naturally leads to a
hierarchy of embedded active length scales, is an appealing
transcription of the hierarchical energy landscape pictures. 
An alternative scenario, based on `temperature chaos', has been actively
promoted in \cite{Yoshino:big}. Although the mechanism leading to rejuvenation 
is different, this scenario also requires a rapid separation of time scales 
as the temperature is reduced to preserve the memory.

\section{Spin glass dynamics, anisotropy and growth of the coherence length}

In order to go deeper into the understanding of the
out-of-equilibrium dynamics of spin glasses and in particular the
rejuvenation and memory effects discussed above, we recently
performed a systematic study of the effect of temperature
variations on the aging dynamics of various spin glasses with
different degrees of anisotropy ranging from Ising to Heisenberg
\cite{Dupuis:PRB2001,Bert:PRL2004}.

Temperature cycling experiments were performed in a
monocrystalline Fe$_{0.5}$Mn$_{0.5}$TiO$_3$ Ising
\cite{Katori:JPSJ1994} sample \#1 ($T_g=20.7K$) and in Heisenberg
SGs with decreasing random anisotropy \cite{Petit:PRB2001}: an
amorphous alloy (Fe$_{0.1}$Ni$_{0.9}$)$_{75}$P$_{16}$B$_6$Al$_3$
\#2 ($T_g=13.4K$), a diluted magnetic alloy Au:Fe$_{8\%}$ \#3
($T_g=23.9K$), an insulating thiospinel
CdCr$_{1.7}$In$_{0.3}$S$_4$ \#4 ($T_g=16.7K$) and the canonical
Ag:Mn$_{2.7\%}$ SG \#5 ($T_g=10.4K$) (data from
\cite{Hammann:PhysicaA1992}).

\begin{figure}
\begin{center}
\includegraphics[width=7.5cm,height=7cm]{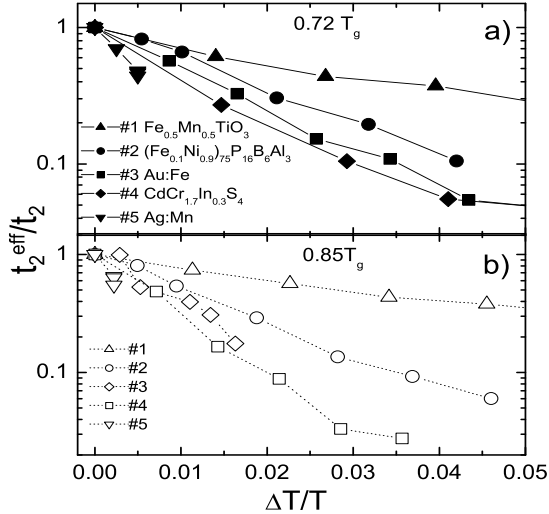}
\caption{\label{fig:4} Effective waiting times deduced from the
temperature cycle experiments at $T \simeq 0.72T_g$ (a, filled
symbols) and $T \simeq 0.85T_g$ (b, open symbols). The SG's are
labelled from \#1 to \#5 by order of decreasing anisotropy.}
\end{center}
\end{figure}

In these experiments, an isothermal TRM procedure allows one to
first define a set of {\it reference curves} that depend on the
waiting time (such as the one in Fig. \ref{fig:1}), and to which
we will compare the relaxation curves obtained after more
complicated histories. In the temperature cycle experiments, the
sample is cooled to a temperature $T<T_g$, kept at $T$ for a time
$t_1=500s$, submitted to long wait at $T-\Delta T$ during
$t_2=9000$s and finally re-heated to $T$ for another short time
$t_3=t_1$. The magnetic field is then switched off and the TRM is
recorded. For each `cycled' TRM curve, an effective waiting time
$t_2^{eff}$ is defined such as to superimpose the obtained TRM curve 
with that of the purely isothermal TRM recorded at temperature $T$,
with waiting time $t_w=t_1+t_2^{eff}(T,\Delta T)+t_3$. For $\Delta
T=0$, the effective waiting time equals the actual waiting time,
but becomes smaller for $\Delta T >0$: $t_2^{eff}(T,\Delta T) <
t_2 $ characterizes the effect at $T$ of aging at $T-\Delta T$.
The results for the different SGs and for two different
temperatures, $T/T_g \simeq 0.72$ and $T/T_g \simeq 0.85$, are
plotted in Fig. \ref{fig:4} (for sample \#5, a similar procedure,
with $t_1=t_3=0$, was applied \cite{Hammann:PhysicaA1992}).

The results of Fig. \ref{fig:4} which correspond to small $\Delta
T$ allow us to probe the separation of length scale discussed in
the previous section \cite{Bouchaud:PRB2002}. Indeed, when considering two temperatures $T$
and $T-\delta T$ such that $\ell^*(T-\delta T)\sim\ell^*(T)$,
rejuvenation and memory effects are no longer observed and aging
tends to accumulate from one temperature to another, which we
described here in terms of an effective time $t_2^{eff}$. The
stronger the decrease of this effective time with increasing
$\delta Ts$, the sharper the separation of active length scales.
Fig. \ref{fig:4} thus shows that the separation of length scales
becomes sharper and sharper when going from the Ising sample
towards Heisenberg spin glasses with weaker and weaker anisotropy.
Our results single out the role of the spin anisotropy in the
nature of the Heisenberg spin glass phase.

A purely activated scenario over constant height barriers would
correspond to a single straight line for all samples in Fig.
\ref{fig:4} \cite{Bert:PRL2004}. In order to account for the
different observed behaviours,  we introduced a crossover expression
for the relationship between time and length scales

\begin{equation}
\label{eq:t(l)}t_w=\tau_0\ell^z exp(\Delta(T)\ell^{\psi}/T)
\end{equation}

\noindent In this expression, which is simply an Arrhenius law
with a temperature-dependent barrier, the usual attempt time $\tau_0$ is
replaced by a renormalized attempt time $\tau_0\ell^z$ where $z$
is the dynamic critical exponent. This modification has the
advantage of giving the expected form for the critical dynamics
close to $T_g$, where due to the vanishing of the barrier height
$\Delta(T)=\Upsilon_0(1-T/T_g)^{\psi\nu}$ the exponential term in
Eq. (\ref{eq:t(l)}) tends to 1. In \cite{Bert:PRL2004} the above
expression was successfully used to explain consistently {\it both} 
temperature
cycling experiments and field variation experiments. This in turn
gave, for each sample, estimates of the barrier exponents $\psi$
and the energy scales $\Upsilon_0$ of the barriers which are consistent with a sharper separation of the
active length scales in Heisenberg-like spin glasses than in the
Ising case (see \cite{Bert:PRL2004} for details). The same trend has recently been observed in large scale simulations of Ising and Heisenberg spin glasses  \cite{BerthierYoung:PRB2004}.

Finally, let us mention an issue that concerns 
the value of the individual spin flip time $\tau_0$ (arising, e.g., in
Eq. (\ref{eq:t(l)})), and that is usually assumed to be microscopic, of the order of
$10^{-12}$ seconds. This assumption is very important since it means, as mentioned
above, that typical experimental cooling rates are extremely slow when compared with $\tau_0$. 
However, it might well be that the flip of a single spin is already slowed down 
by its environment (interactions, local anisotropy...). 
In the Ising case, this slowing down might be somewhat large, and could significantly 
renormalise the value of $\tau_0$, bringing it much closer (especially at low
temperatures) to the experimental time window. This could significantly affect 
the precise values of the parameters involved in Eq. (\ref{eq:t(l)}).

\section{Conclusion}

In this paper, we have briefly reviewed several important features of the slow
out-of-equilibrium dynamics of spin glasses. We have first
discussed the scaling laws that govern the isothermal aging of the
response function of spin glasses, and showed experiments which
suggest that the observed sub-aging behaviour may be intrinsic, although 
when compared to a similar effect observed in numerical simulations,
laboratory cooling rates might still be too slow to conclude. We
have then discussed the interpretation of the rejuvenation and
memory effects observed in spin glasses submitted to temperature
variations during their aging. In both energy landscape and
real space pictures, the need for a hierarchy of state/length
scales seems unavoidable for the description of such effects. This 
strong separation of length scales probed at different temperatures 
is also needed in the `temperature chaos' interpretation of these 
rejuvenation/memory effects (see \cite{Yoshino:big} for a review and 
detailed arguments supporting this point of view, and \cite{Martin-Mayor:2004} 
for very recent numerical simulations challenging this interpretation). 

We have reported experiments probing the effect of very
small temperature variations in spin glass samples with various
degrees of anisotropy. The main results of this last investigation
are (a) a clear finger-print of a dynamical crossover between a critical 
regime and an activated regime in spin-glasses, with energy barriers vanishing 
at $T_c$ and (b) a trend to a sharper separation of 
active length scales with temperature in Heisenberg-like samples.

\vskip 1cm

This work was supported in part by the European Community's Human
Potential Programme (contract HPRN-CT-2002-00307 referred to as
DYGLAGEMEM).

\end{document}